\documentclass[conference]{IEEEtran}
\usepackage{tikz}

\usepackage{ifpdf}

\usepackage{cite}

\usepackage[]{graphicx}
\usepackage[hyphens]{url}
\usepackage[pdfpagelabels]{hyperref}
\usepackage[hyphenbreaks]{breakurl}
\usepackage{float}
\usepackage{epstopdf}
\DeclareGraphicsExtensions{.eps,.pdf,svg}

\ifCLASSINFOpdf
\else
\fi
\usepackage{mathtools}
\usepackage{url}
\usepackage{threeparttable}
\usepackage{tablefootnote}

\begin{document}

%

\title{{\huge Performance Analysis and Comparison of\\ Distributed Machine Learning Systems}}


\author{
Salem Alqahtani\\
University at Buffalo, SUNY\\
salemmoh@buffalo.edu
\and
Murat Demirbas\\
University at Buffalo, SUNY\\
Demirbas@buffalo.edu
}


%


\maketitle

\begin{abstract}

Deep learning has permeated through many aspects of computing/processing systems in recent years. While distributed training architectures/frameworks are adopted for training large deep learning models quickly, there has not been a systematic study of the communication bottlenecks of these architectures and their effects on the computation cycle time and scalability. In order to analyze this problem for synchronous Stochastic Gradient Descent (SGD) training of deep learning models, we developed a performance model of computation time and communication latency under three different system architectures: Parameter Server (PS), peer-to-peer (P2P), and Ring allreduce (RA). To complement and corroborate our analytical models with quantitative results, we evaluated the computation and communication performance of these system architectures of the systems via experiments performed with Tensorflow and Horovod frameworks. 

We found that the system architecture has a very significant effect on the performance of training. RA-based systems achieve scalable performance as they successfully decouple network usage from the number of workers in the system. In contrast, 1PS systems suffer from low performance due to network congestion at the parameter server side. While P2P systems fare better than 1PS systems, they still suffer from significant network bottleneck. Finally, RA systems also excel by virtue of overlapping computation time and communication time, which PS and P2P architectures fail to achieve.  

\end{abstract}

\IEEEpeerreviewmaketitle

\section{Introduction}

Deep Neural Networks (DNNs) have dramatically improved the state-of-the-art for many problems that machine learning (ML) and artificial intelligence (A.I) community have dealt with for decades, including speech recognition, machine translation, object identification, self-driving cars, and healthcare record analytics and diagnostics. DNN training is hungry for big data and high computation power, and even high-end machines are inadequate to respond to this demand\cite{memory}. Thus, distributed DNN training architectures/ frameworks that utilize a cluster of computers have quickly become popular in recent years~\cite{abadi2016tensorflow,sergeev2018horovod}.
As these distributed training frameworks need to coordinate the nodes in the cluster efficiently for sharing states, parameters, and gradients, they are confronted with many challenges in terms of consistency, fault tolerance, communication overhead, and resource management~\cite{challengings}. 

Three distributed training architectures have emerged for DNN training. Parameter server (PS) architecture uses a number of parameter servers that serve to coordinate/ synchronize model updates by a number of workers.
The workers pull the model from the parameter-servers, compute on the DNN, and then send the computed gradients to the parameter servers. In the peer-to-peer (P2P) model, worker and server processes coexist on the same machine. The worker process pulls the model locally from the server process in the same machine, computes on the DNN and sends the computed gradients to every other machine. Finally, in the Ring allreduce (RA) model, there is only server process on every machine. The server reads the model from its buffer and computes on the DNN, and sends the computed gradients to its neighbor in the ring.

The developers need to choose among these architectures and configure the framework with the number of workers and servers, depending on the workload and available network and computing infrastructure. While there has been a lot anecdotal evidence that different architectures and different configurations that lead to drastically different performance~\cite{ben2018demystifying}, there has not been a systematic study of communication bottlenecks of these architectures and their effects on the performance of training.

In this paper, we investigate this problem. Since synchronous Stochastic Gradient Descent (SGD) works best for DNN training~\cite{chen2016revisiting}, we make it our focus. We take a two-pronged approach and investigate the three architectures both analytically and empirically.

For analytical assessment of the PS, P2P, and RA architectures, we develop models for latency (total time for training one epoch), which includes computing time, and communication time. The computing time is the time spent to compute the DNN, and the communication time is the time spent to send the training result to a server or servers. Knowing both times is essential to understand the behavior of these systems, and to optimize the overlapping period between both times. Our model analysis shows that the dominant part is often the communication rather than the computation time during training process, and is able to rank the network use/congestion of the three architectures modeled.

To complement and corroborate our analytical models with quantitative results, we evaluate the computation and communication performance of these systems via experiments performed with Tensorflow and Horovod frameworks. To measure the convergence speed, we provide a quantitative evaluation. We perform experiments with PS, P2P, and RA architecture and compare it to our model results. More specifically, we measure the throughput (amount of training samples per second) and latency for large-scale ML systems with TensorFlow system~\cite{abadi2016tensorflow} and Horovod system~\cite{sergeev2018horovod}. We choose TensorFlow to take advantage of the high usability and high abstraction level for operations and devices. We use Horovod library to take advantage of MPI features and its integration with TensorFlow. The dataset we feed to our models is the MNIST handwritten digits which is widely used in research.



Our results show that RA achieves high throughput and low latency compared to PS and P2P systems. This is because, in RA the available network bandwidth is constant between worker nodes whereas for the PS or P2P systems, the bandwidth is a shared resources among all worker nodes.  We also find that the RA system achieves a high overlapping between computation time and communication time than PS and P2P systems. Finally, we find that P2P and PS systems have load imbalance among peers because tensor size in each DNN layer is different.

\vspace{3mm}
\noindent
{\bf Outline of the rest of the paper.} We give background on DNNs in Section~\ref{sec:back}. In Section~\ref{sec:permodel}, we develop a performance models for distributed training for PS, P2P, and RA architectures. We evaluate the performance of these three architectures in Section~\ref{sec:eval}. In Section~\ref{sec:rw}, we summarize related work and conclude the paper in Section ~\ref{sec:concl}.

\section{BACKGROUND}
\label{sec:back}

In this section, we explain the neural network training process on a single computer node, and then describe distributed neural network training on multiple nodes.

\subsection{Artificial Neural Network}
\label{sec:DDNNs}

\begin{figure}[htb]
\includegraphics[width=3.2in]{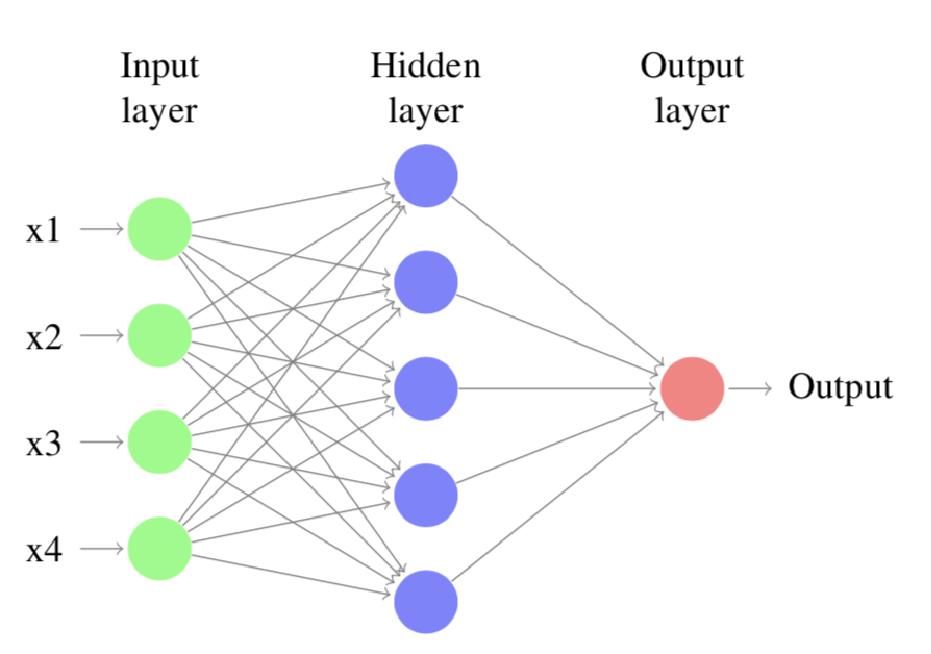}
\caption{Deep Neural Networks}
\label{fig:DDNNs}
\end{figure}

Artificial neural networks are computing systems for processing complex data input for many ML algorithms~\cite{wiki:ANN}. Here, our focus will be on multi-layer Neural networks shown in Figure~\ref{fig:DDNNs}, which is a set of connected input/output to computation units where each connection has a weight associated with it. During the training phase, the network learns by adjusting the weights to be able to predict the correct class label of the input samples. The basic neural network architecture categorized as an input layer, hidden layers, and an output layer. The input layer reads input data instances $X$, while the output layer holds and displays the results of the neural network. Each set of neurons is grouped in a single layer. A single neuron represents computational unit for input and weight values that neuron receives from a previous layer. The modern deep neural networks architectures aim to train on very large datasets with huge parameters in order to improve the performance for many real-world applications.

To compute DNNs, the first step is feeding the network with weighted edges $W$ and a dataset $D=(X,Y) $ where $X$ represents the data instances and $Y$ represents labels. This step is formally called feedforward neural network, where the data move from the previous layer $(L-1)$ to the next $(L)$ layer forming no cycle. The $a^{(0)}$ denotes the input data $X$ at the input layer. Then, neuron will sum the products of its input, weighted values, and bias term $b$. 

\begin{equation}
Z^{(L)} = ( W^ {L} * a^{(L-1)} +b^{(L)})
\end{equation}

Every neuron has an activation function for instance $\sigma(.)$. The total value will pass through a non-linear activation function. If the total value is above the threshold, the neuron will fire, otherwise it will not. 

\begin{equation}
a^{(L)} = \sigma(Z^{(L)})
\end{equation}

Then, we compare the predicted output value $a^{(L)}$ with the ground truth $Y$ in the training data $D$, and measure the error using a loss function~\cite{Loss}. The loss function is an objective function that should be minimized until the model converges.

\begin{equation}
C = ( a^{(L)} - Y )^2
\end{equation}

After calculating feedforward pass and loss function, the neural networks use a back-propagation algorithm~\cite{rumelhart:errorpropnonote} to train neural network model values. The back-propagation computes the gradients by propagating the error (the difference between the targeted and actual output values) to every individual neuron.

\begin{equation}
\frac{\partial{C}}{\partial{W^{L}}} = \frac{\partial{Z}^{(L)}}{\partial{W^{(L)}}} * \frac{\partial{a^{(L)}}}{\partial{Z^{(L)}}} *  \frac{\partial{C}}{\partial{a^{(L)}}}
\end{equation}

\subsection{Distributed Neural Network Training}

In recent years, the advance of hardware, training methods, and network architectures have enabled distributed training which minimizes the training time for DNNs training. Instead of restricted to a single machine, now we can scale to as many resources as required. In this paper, we choose to perform on data parallel distributed training, but we also explain the model parallel distributed training. Many practitioners avoid using model parallelism because of the network overhead that it creates, due to layers distribution on many machines. Finally, the use of data parallelism models that learn over large amounts of training data are more common than models with billions of parameters~\cite{li2015malt}. 

\subsubsection{Model Parallelism}
\label{sec:mp}

\begin{figure}[htb]
  \includegraphics[width=3.2in]{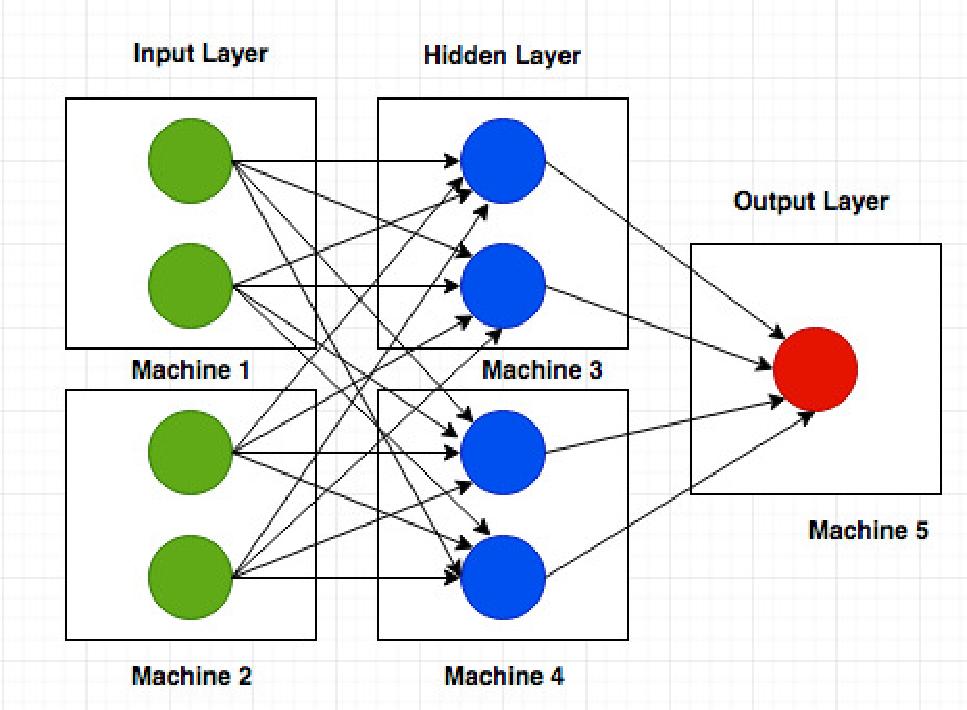}
  \caption{Model Parallelism.}
  \label{fig:Model_Para}
\end{figure}

The model parallelism scheme is shown in Figure~\ref{fig:Model_Para}. The parallelization mechanism here is to split the model parameters $W$ and $b$ among many nodes. Each node is responsible for doing some computation tasks in a different part of the network. The node will communicate the neurons activities with other machines after finishing local computation. The limitations for the model partition training are difficult to partition the models because each model has its own characteristics and high communication latency between devices. However, it is rarely used in the real world applications because of the challenges to get good performance, but it is preferable when a node is not sufficient to store all the model parameters and computationally expensive.

\begin{figure}[htb]
  \includegraphics[width=3.2in]{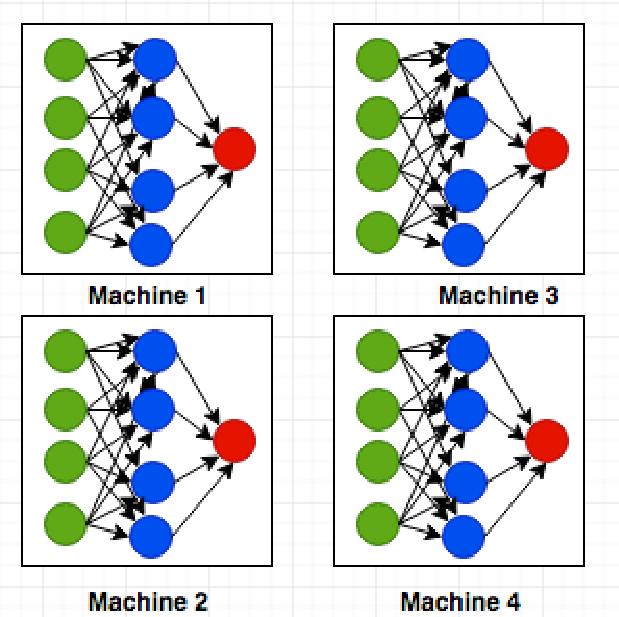}
  \caption{Data Parallelism.}
  \label{fig:Data_Para}
\end{figure}

\subsubsection{Data Parallelism}
\label{sec:dp}

In the data parallelism scheme, as shown in Figure~\ref{fig:Data_Para}, each worker machine creates a complete computation graph and typically communicates gradients $\Delta W$ with a model parameter holder such as PS. The data parallelism scheme is used extensively in many applications due to its simplicity. The data samples are partitioned and assigned across all computation nodes (eg. GPU,TPU) $\frac{n}{w}$. This is the contrast with the model parallelism, which uses the same data for every worker machine but partitions the model among the worker machines. Each node computes independently from every other node has a subset of the dataset and synchronizes computation results. Mini-Batch SGD is a common approach and shows great performance in many models. Mini-Batch SGD makes the update on a subset of the dataset at each iteration rather than using an entire dataset at each iteration. Each worker trains on different data samples, and exchanges different outputs by network communication with other replicas in the system to update the model until it reaches consensus. Data parallel adjusts the weight values $W$ using widely used method called gradient descent algorithm~\cite{SGD} for combining results and synchronizing the model parameters between each worker.

\begin{equation}
W_{i+1} := W_{i} - \alpha\frac{\partial{C}}{\partial{W_{i}^{L}}}
\end{equation}

\begin{equation}
b_{i+1} := b_{i} - \alpha\frac{\partial{C}}{\partial{b_{i}^{L}}}
\end{equation}

The data parallelization merits are to increase the system throughput through distributed parallel computing and to handle  exchange high data volum. However, this approach is limited by the available optimization algorithms and hardware.

\subsubsection{Bulk Synchronous Parallel}
\label{sec:bsp}

In distributed computing systems, each computing node has different computing power than other nodes in the system due to real-world environment. For this reason, the distributed ML training uses an iteration to coordinate the synchronization between all computer nodes~\cite{zhang2017parameter}. In the synchronous update known as bulk synchronous parallel (BSP)~\cite{BSP}, the replicas submit the gradients after locally training process at every iteration or mini-batch to global model parameters or to other replicas. Then, each node stops by a synchronization barrier from training the next iteration until global model receives all results of other active workers. The downside of this approach is that the training time will be dominated by the strugglers and each iteration requires a lot of communication. Also, the workers must enter the synchronization barrier which takes a non-trivial amount of time to exit the synchronization barrier~\cite{xing2016strategies}. However, in a synchronous approach, the algorithm converges relatively faster than asynchronous training. The reason is that there is no stale gradient because in each iteration the gradients collect from all replicas and the model update in the next iteration. 

\subsubsection{Stale Synchronous Parallel}
\label{sec:ssp}

In the stale synchronous parallel (SSP)~\cite{ho2013more}, the replicas execute their local iterations and go to the next iteration without a synchronization barrier. When the faster machines exceed the slower machines by S iterations which is a threshold, all nodes enter synchronization barrier allowing other machines to be synchronized. All gradients in a given mini-batch are computed and sent to the global parameter model. Then, replicas pull new model parameters with stale gradients before all others sent their update from previous iterations. For example, with $N$ replicas, the gradients for some replicas calculated from the stale parameters copies related to previous iterations. The global model parameters update is not more than bounded iterations to reduce the synchronization overhead. The interleaving computation with communication is the greatest benefit for using SSP communications. The algorithm also has a slow convergence rate.

\section{Performance Modeling}
\label{sec:permodel}

To model the behavior of the system, and to estimate the system performance, we present a performance model that captures the computation time and communication latency based on varying system configurations. We deal with the systems at a high abstraction layer due to the complexity of the systems. Large-scale platforms have different underlining designs~\cite{KUO1} such as TensorFlow, and Spark, and for that, it becomes difficult to design a performance model at a low level. Our model approximates both computation and communication runtime for a single epoch (a single pass through the full training set) of training DNN with mini-batch SGD. In this work, we present a performance model that is simple and accurate enough to calculate computation time and communication latency without intensive log data collection. 

Our results have two network indicators, latency (the time it takes to send a message from point A to point B) and throughput (the processed amount of data every time unit). These indicators differ from one system design to another. Modeling network latency has two factors. Download time for workers receive data from the server while upload time for workers send gradients to servers. For further details, the parameters of the performance model PS, P2P, and RA in Table \ref{table:1}.

\begin{table}[h!]
\centering
\begin{tabular}{c c} 
\hline\hline 
Notation & Meaning \\ 
\hline\hline
\multicolumn{2}{c}{Machine Learning Notation} \\
\hline
$W^{L}$ & weight variables at layer L\\
$b^{L}_{i}$ & bias variables i at layer L \\
$a^{(L)}$ & forward Pass\\
$\sigma(.)$ & activation function \\
$Y$ & ground truth \\
$C$ & loss function \\
$n$ & Number of training examples\\
$\alpha$ & learning rate \\
$b$ & batch size\\
$t_{i}$ & iteration number i \\
m & mini-batch size \\
W & model size \\
\hline\hline
\multicolumn{2}{c}{Distributed Systems Notation} \\
\hline\hline
PS & Parameter Servers\\
w & number of worker \\
B & total bandwidth \\
\hline
\end{tabular}
\caption{Performance mode notation table}
\label{table:1}
\end{table}

\subsection{Distributed Training with PS System}
\label{sec:permodelcs}

\begin{figure}[htb]
  \includegraphics[width=3.2in]{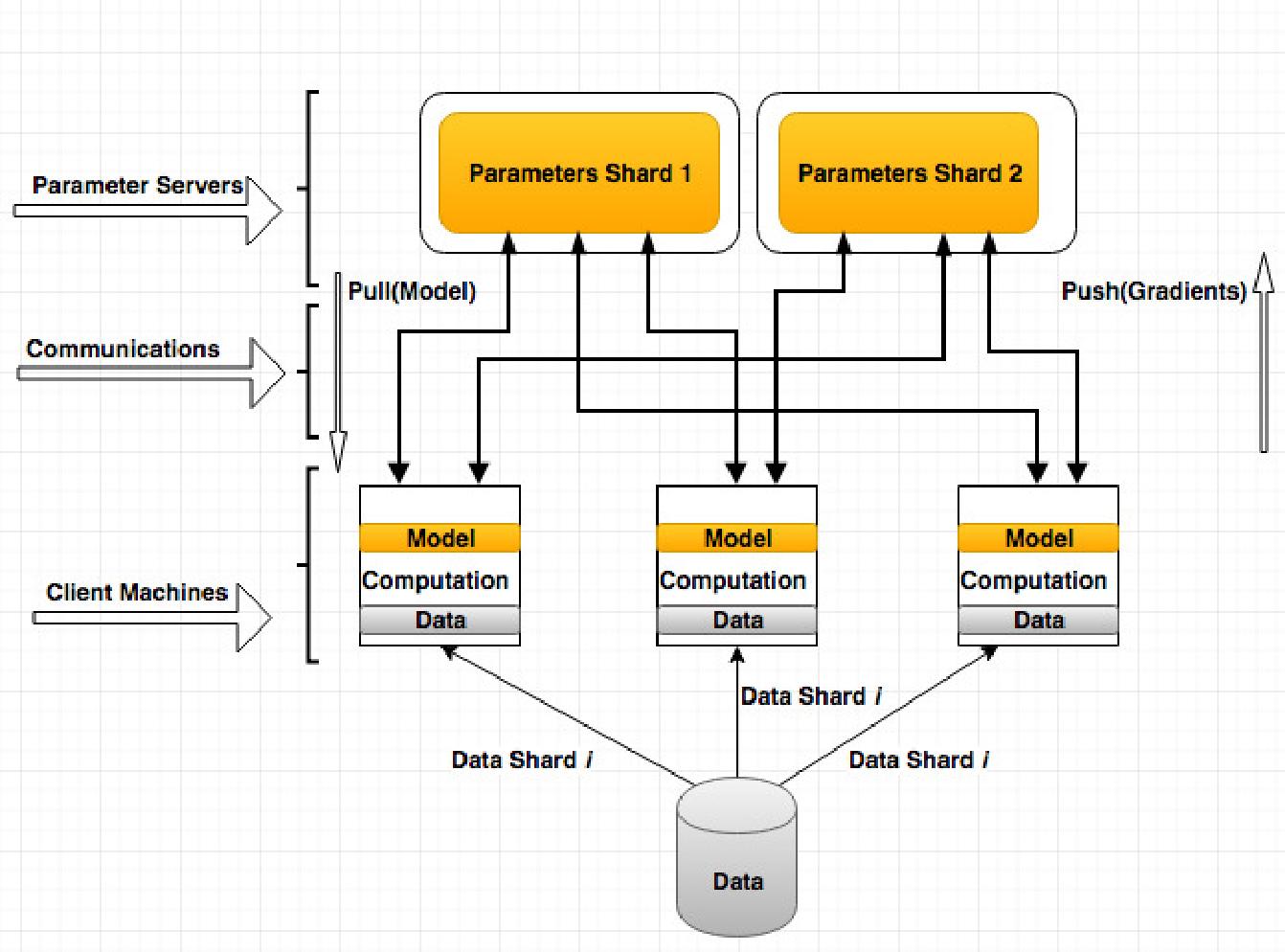}
  \caption{PS Architecture.}
  \label{fig:ps}
\end{figure}

The PS as shown in Figure~\ref{fig:ps} was introduced in~\cite{smola2010architecture}, and was followed with second, and third generations~\cite{dean2012large, li2014scaling}. The PS system was built to solve distributed ML elastic scalability, communication, and flexible consistency problems~\cite{li2014communication}. It is a key-value store dedicated storing variables and does not conduct any computation task. The PS setup might consist of 1PS node or many PS nodes, each of which maintains a subset of ML parameters (weights and bias). The PS adapts one-to-all, and all-to-one collective communication topology for exchanging the gradients and model between servers and workers using different mechanisms such as Google gRPC protocol, a default communication protocol of TensorFlow framework based on TCP~\cite{website2013}, as illustrated in Figure~\ref{fig:ps}. 

Often, the PS system uses parallelism technique called data parallelism, as described in~\ref{sec:mp} where training dataset splits into small batches called mini-batches that are used to calculate model error and update model parameters. 

Dataset and workload are divided equally among all active worker nodes in the system. The PS starts by broadcasting the model to the workers. Each worker performs a neural network computations by reading its own split from the mini-batch, and computing its own gradients. Workers communicate with all PSs to send their training results. The PS incorporates gradients from all nodes and updates the stored model parameters~\cite{iandola2016firecaffe}. Many similar systems of PS has adapted key-value store interfaces~\cite{dai2013petuum}. The worker nodes use key-value store API to pull the recent parameters from the PS (i.g. pull()), and to push the gradients to the PS (i.g. push()). The PS extends the single server to more than one to solve load balance, and to reduce communication bottlenecks.  

\begin{equation}
T_{(processing)} = (T_{a{(L)}} + T_{(C)} + T_{(\frac{\partial{C}}{\partial{W^{L}}})})
\end{equation}

The above formula shows the computation time for training DNNs including feedforward time, loss function time, and back propagation time. Most well-know ML framework systems who adapted this architecture design are TensorFlow from Google, MXNet from Amazon. 

The PS training is proceeding at the speed of the slowest machine in any iteration with a synchronous model while an asynchronous model overcome strugglers node who degrade the training speed but  asynchronous model may affect the general model accuracy. To choose between these communication models, developers and researchers trade-off accuracy and speed which depends on their applications~\cite{cui2014exploiting}. The downside of the PS becomes a communication bottleneck which leads to slow down the training process and to limit the system scalability in very large-scale. Removing the central server bottleneck in asynchronous distributed learning systems while maintaining the best possible convergence rate is the optimal design solution~\cite{cui2014exploiting, cui2014big, cheung2002effect, ahmed2012scalable}. The Formula~\ref{eq:B1} shows how much bandwidth is dedicated for each worker node to communicate with the PS.

\begin{equation} \label{eq:B1}
\begin{split}
available_{B} =((\frac{1}{w} * {Total Bandwidth}) * ps )
\end{split}
\end{equation}

Notice that the $available_ {B}$ is the available bandwidth between every client and PS, $Total Bandwidth$ represents the total bandwidth available for all clients, and $w$ indicates all active workers who communicate with PS. $w\geq ps$, workers should not have less than the number of PS nodes in order to divide the bandwidth evenly among workers.

\begin{equation} \label{eq:data}
\begin{split}
Pull(W) = (epoch * \frac{W}{available_{B}} * (\frac{\frac{n}{b}}{w}))
\end{split}
\end{equation}

Where $Pull(.)$ is defined as a delay function for pulling weight values through the communication link. The $W$ is the model size. $\frac{n}{b}$ defined the number of times that workers pull the model from PSs in a single epoch. The workers compute the gradients and push the results to the PS that aggregates the gradients after the majority of the nodes communicate their gradients and a new result will be pulled from the workers for the next iteration. 
\begin{equation} \label{eq:data1}
\begin{split}
Push(W) = (epoch * \frac{(\frac{W}{w})}{available_{B}} * (\frac{\frac{n}{b}}{w}))
\end{split}
\end{equation}

Most frameworks like TensorFlow and MXNet, parallelize the gradient aggregation of the current layer of the neural network with the gradient computation with the previous layer. This optimization hides the gradient communication overhead. In above formula, we calculate the time that takes workers to push the gradients to the PS, where $\frac{W}{w}$ model size divided by the number of workers which is our formula to calculate the gradients size.

\begin{equation}
T_{(cpu-ps)} = (epoch* (\frac{\frac{n}{b}}{w}) * (T_{processing}+ PS\_time))
\end{equation}

\begin{equation}
T_{(tcp-ps)}  = Pull(W) + Push(W)
\end{equation}

\begin{equation}
T_{(total)}  = T_{(cpu-ps)}  + T_{(tcp-ps)} 
\end{equation}

\begin{figure}[htb]
  \includegraphics[width=3.2in]{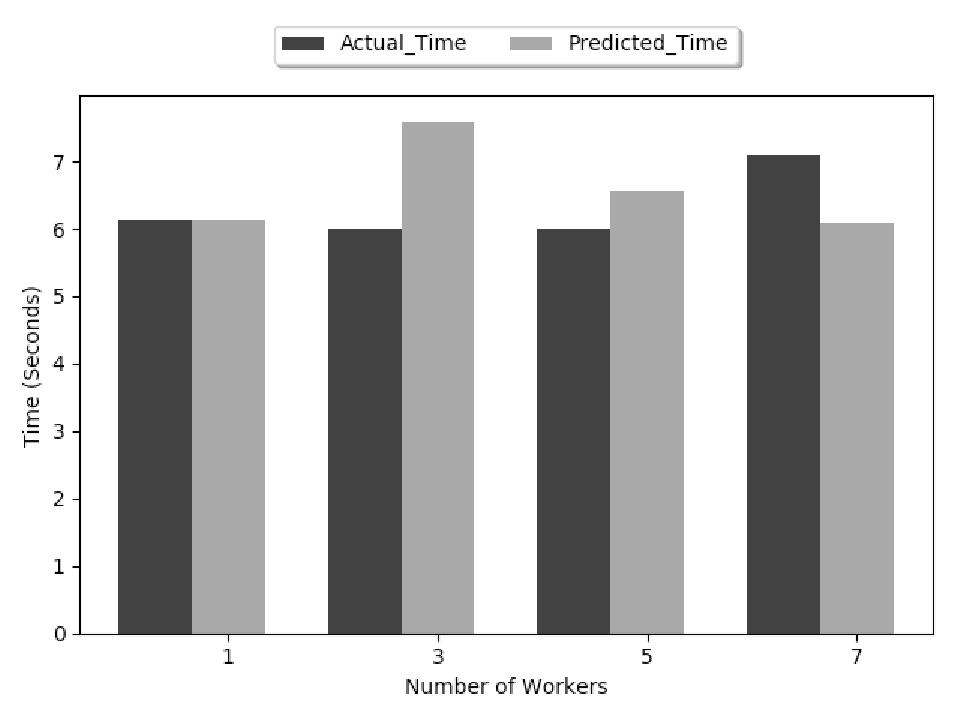}
  \caption{Estimated epoch time for 1PS.}
  \label{fig:psmodel}
\end{figure}

\begin{figure}[htb]
  \includegraphics[width=3.2in]{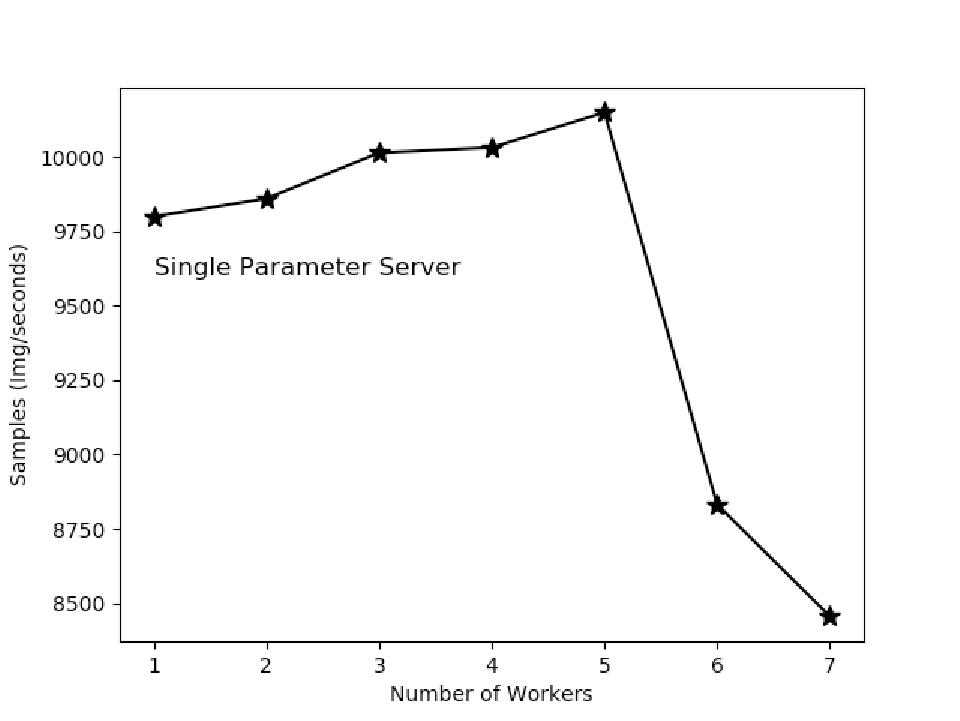}
  \caption{Measured training throughput of 1PS.}
  \label{fig:psthroughput}
\end{figure}

\begin{figure}[htb]
  \includegraphics[width=3.2in]{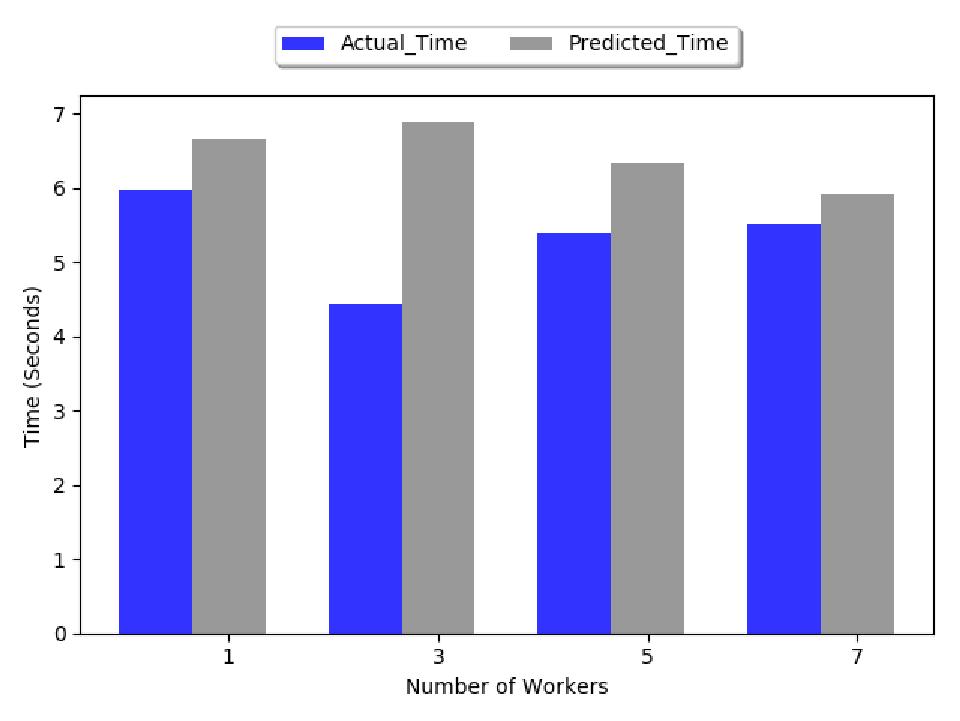}
  \caption{Estimated epoch time for 2PS.}
  \label{fig:2psmodel}
\end{figure}

\begin{figure}[htb]
  \includegraphics[width=3.2in]{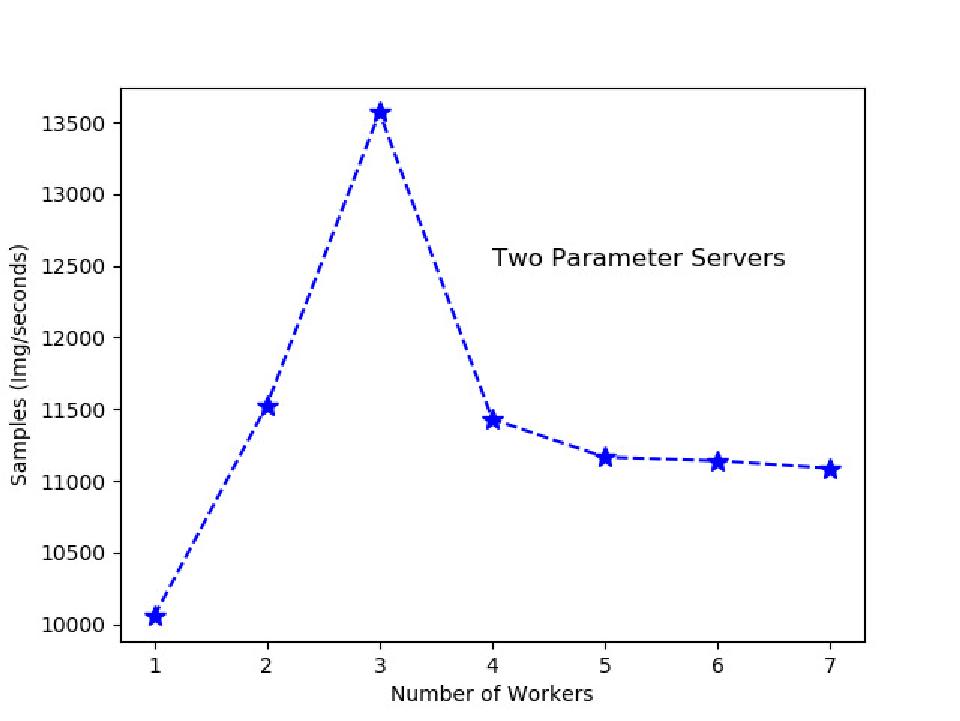}
  \caption{Measured training throughput of 2PS.}
  \label{fig:2psthroughput}
\end{figure}

The above formulas for $T_{(cpu-ps)}$ calculate approximate computation time for computing neural network which is a learning variable defined and extracted from experiments because each dataset has a different number of features that lead to different computation time cost. $PS\_time$ is a time that PS takes to update the model with new results from workers. This time is a constant time in our analysis because the number of features in datasets is variant. The communication runtime $T_{(tcp-ps)} $ in this model has linear complexity and $Push(W)$ time often is overlapped under computation time. Basically, we have three basic blocks, computing time, communication time, and synchronization time. Our performance model for capturing a single PS $T_{(total)}$ comparing to actual runtime is shown in Figure~\ref{fig:psmodel}. If there is more than one PS, each one should maintain a portion of global shared parameters and communicates with each other to replicate and to migrate parameters for reliability and scaling. From $T_{(total)}$, the expected runtime for 2PS is shown in Figure~\ref{fig:2psmodel}. We show the system throughput for one and two PSs in the Figures~\ref{fig:psthroughput}, and~\ref{fig:2psthroughput}. The ideal throughput in distributed training increases linearly with number of worker nodes.

\begin{figure}[htb]
  \includegraphics[width=3.2in]{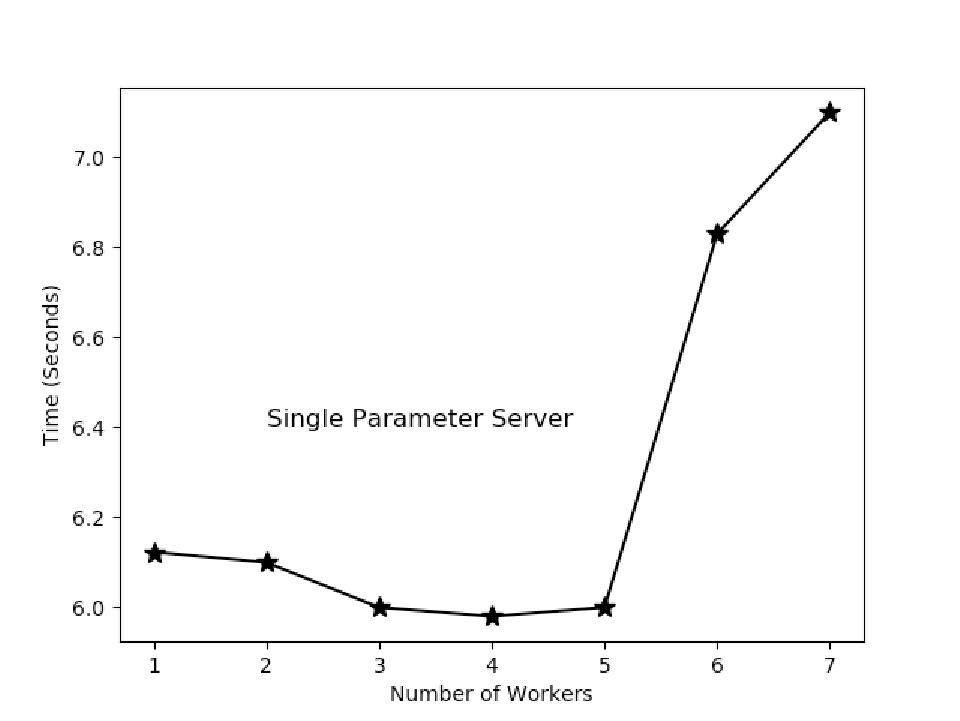}
  \caption{Epoch time for 1PS.}
  \label{fig:pslatency}
\end{figure}

\begin{figure}[htb]
  \includegraphics[width=3.2in]{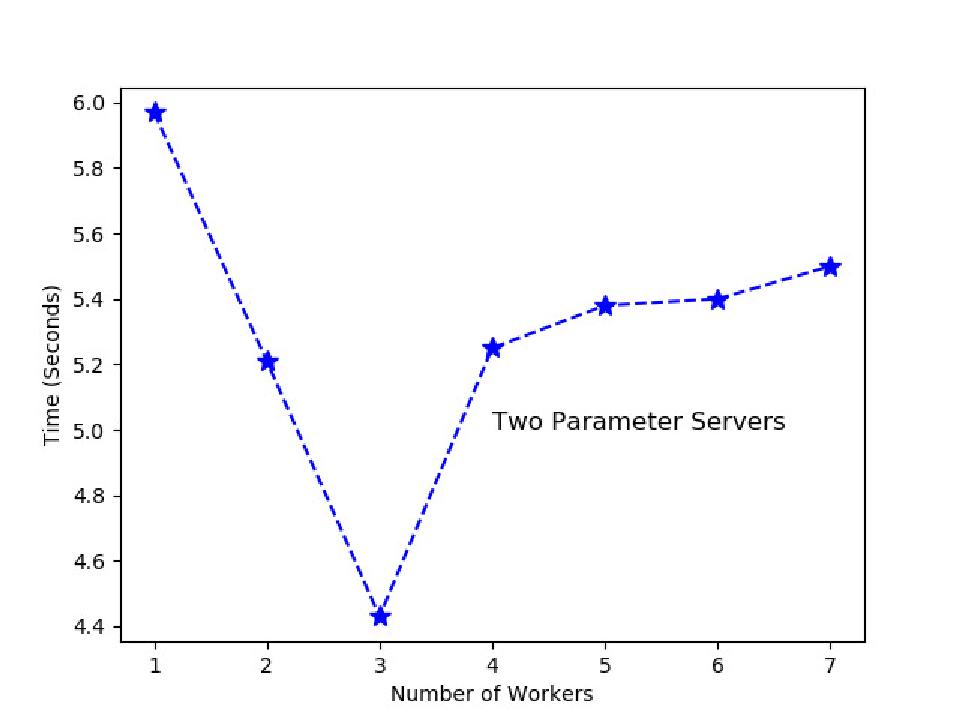}
  \caption{Epoch time for 2PS.}
  \label{fig:2pslatency}
\end{figure}

The formula below calculates the ideal samples per second with respect to number of workers. 

\begin{equation}
Ideal = (T_{single} * n) * w
\end{equation}

\begin{figure}[htb]
  \includegraphics[width=3.2in]{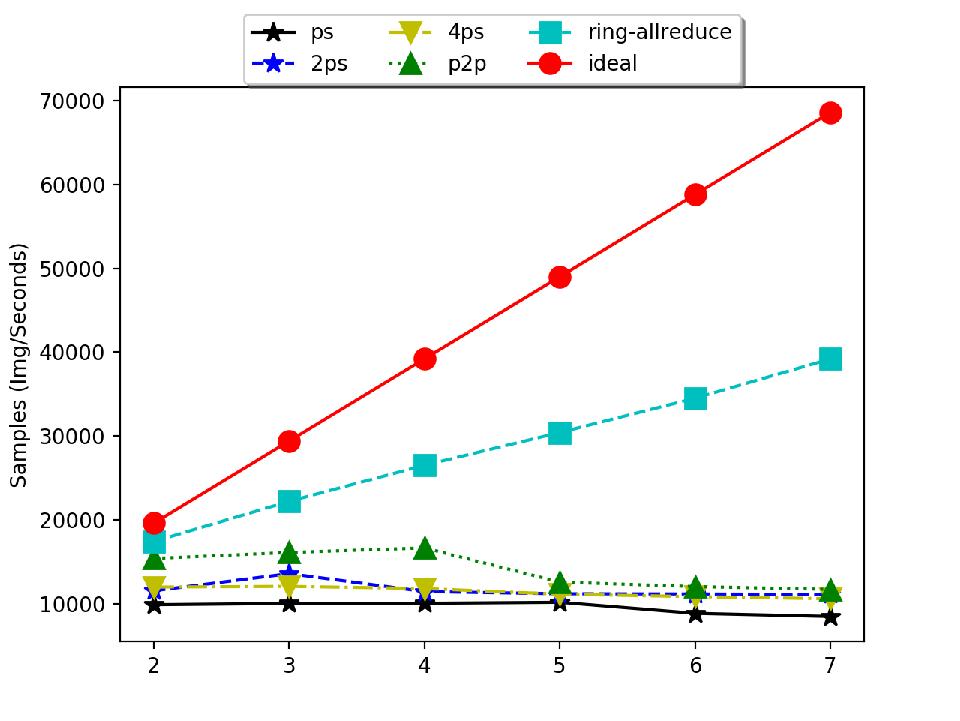}
  \caption{Comparison of Ideal Throughput with Actual System Throughput for RA, P2P, and PS.}
  \label{fig:ideal}
\end{figure}

In Figure~\ref{fig:ideal}, we compare the ideal samples per seconds with actual system throughput based on our experiment. The $T_{single}$ denotes for the single training processing time. 

\subsection{Distributed Training with P2P System}
\label{sec:permodelp2p}

\begin{figure}[htb]
  \includegraphics[width=3.2in]{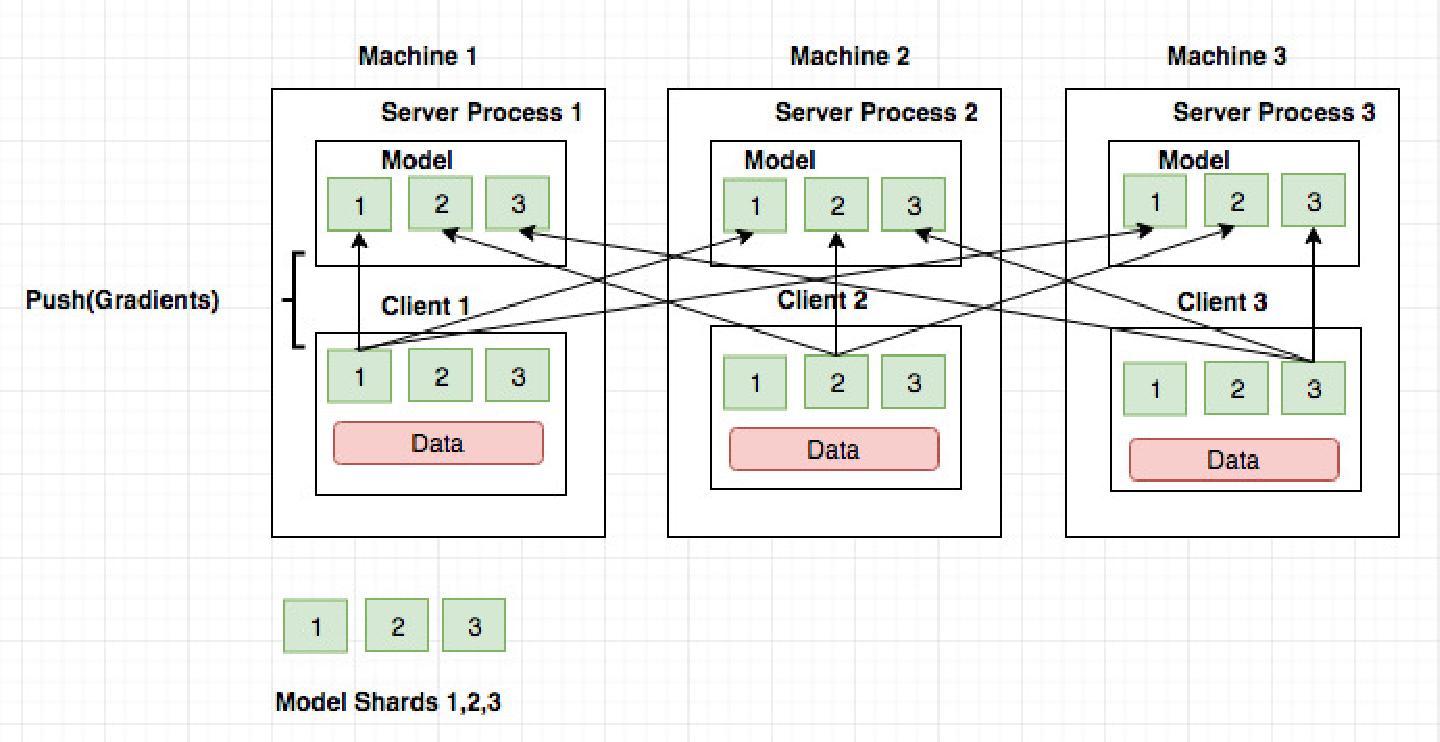}
  \caption{P2P based Architecture.}
  \label{fig:p2p}
\end{figure}

\begin{figure}[htb]
  \includegraphics[width=3.2in]{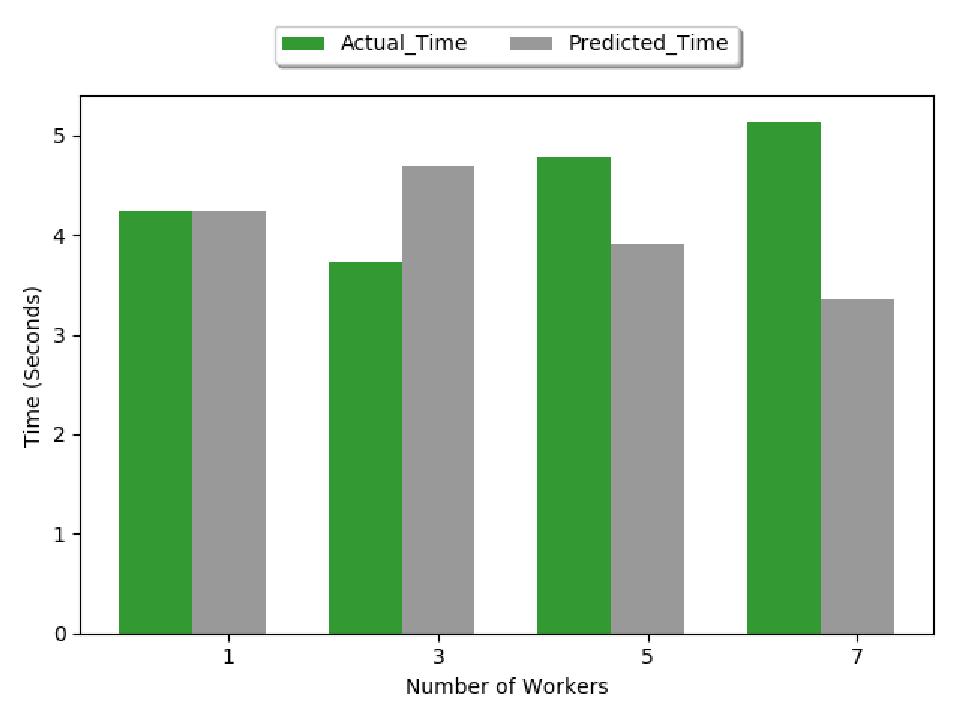}
  \caption{Estimated epoch time for P2P system.}
  \label{fig:p2pmodel}
\end{figure}

\begin{figure}[htb]
  \includegraphics[width=3.2in]{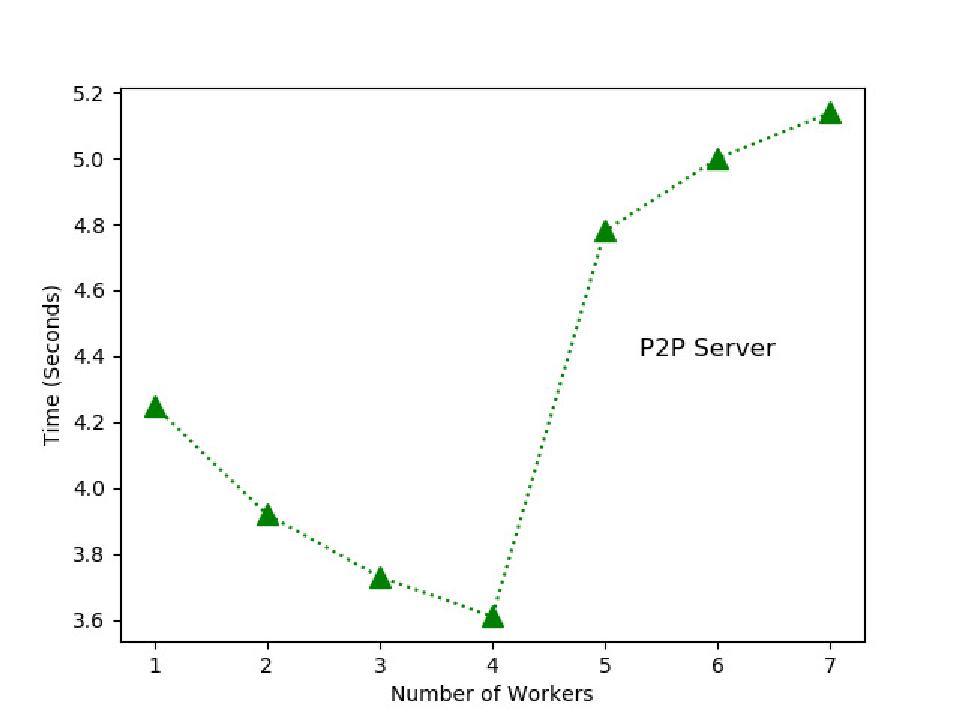}
  \caption{Epoch time of P2P system.}
  \label{fig:p2platency}
\end{figure}

\begin{figure}[htb]
  \includegraphics[width=3.2in]{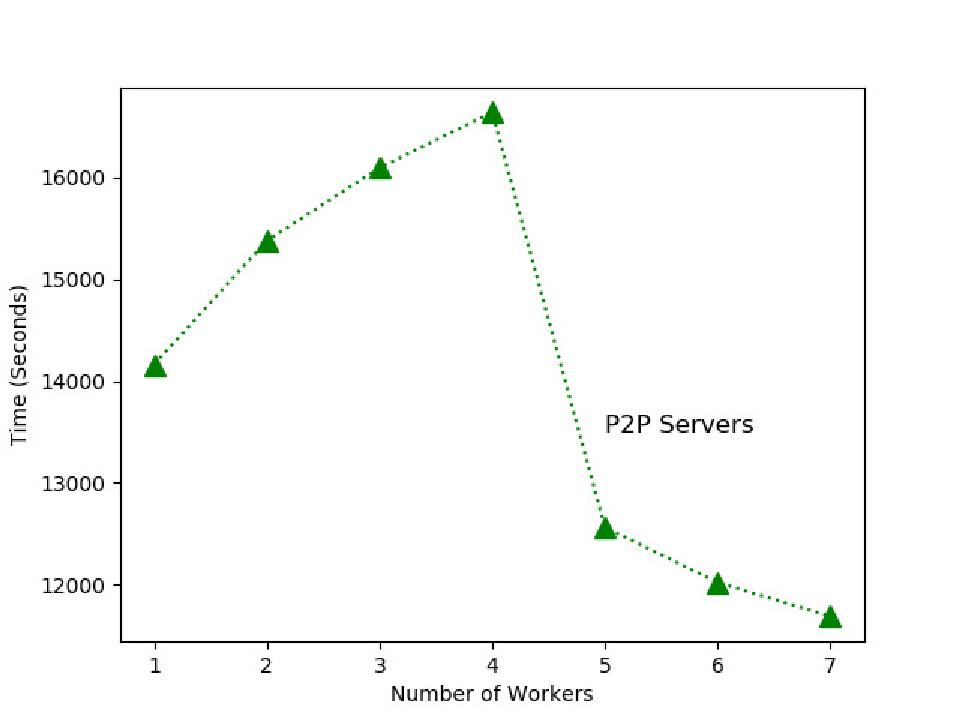}
  \caption{Measured training throughput of P2P system.}
  \label{fig:p2pthroughput}
\end{figure}

In this system design, as shown in Figure~\ref{fig:p2p}, every node joins the system is a peer. Peers connect to one another and provide the functionality of saving model parameters and training the neural networks. The first P2P data-parallel system library to solve large-scale ML problems was introduced in~\cite{li2015malt}. At the high level, both client and server reside on the same machine, which allows replicas to send model updates to one-another instead of a central PS, as shown in Figure~\ref{fig:p2p}. Initially, all nodes obtain the same model and subset of the dataset. Each client node calculates feedforward and back propagation passes over mini-batch SGD. At the end of each iteration, the workers push a subset of model parameter updates $\Delta W$ to parallel model replicas to ensure that each model receives the most recent updates from other nodes. The total epoch time that consists of communication delay of sending gradients for all other peers, receiving model from same machine, and computation time is shown in figure~\ref{fig:p2platency}. In figure~\ref{fig:p2pthroughput}, we show the total system throughput that nodes have processed every time unit. In Figure~\ref{fig:p2pmodel}, we compare the performance model with the actual running time. One advantage of this P2P model software simplicity because the developers write only one code and distributed on all active machines. However, this approach is limited by optimization algorithms and available hardware.
Reading model parameters size and writing gradients size are different because workers read whole model size from the same machine while in writing the workers update subset of the model size through network. Recently, most DNN frameworks overlap computation time with gradients updates.

\begin{equation}
T_{(cpu-p2p)} = (epoch* (\frac{\frac{n}{b}}{w}) * (T_{processing}+ update\_time))
\end{equation}

\begin{equation} \label{eq:B}
available_{B} =(\frac{Total Bandwidth}{2*(w-1)})
\end{equation}

\begin{equation}
T_{(tcp-p2p)}  =  (\frac{\frac{W}{w}}{available_{B}} * (\frac{(\frac{n}{b})}{w}))
\end{equation}

\begin{equation}
T_{(total)} = T_{(cpu-p2p)} + T_{(tcp-p2p)}
\end{equation}

Here, we are not interested in one iteration time, but we are interested in an epoch time. $available_{B}$ is the bandwidth between the peer and other peers. In this model, we noticed that gradients have not perfectly overlapped with computation time and have high overhead. In every iteration, server will send and receive $2*(w-1)$ messages. The message size will be equals to $(\frac{W}{w})$.

\subsection{Distributed Training with Ring-allreduce}
\label{sec:permodelr}

\begin{figure}[htb]
  \includegraphics[width=3.2in]{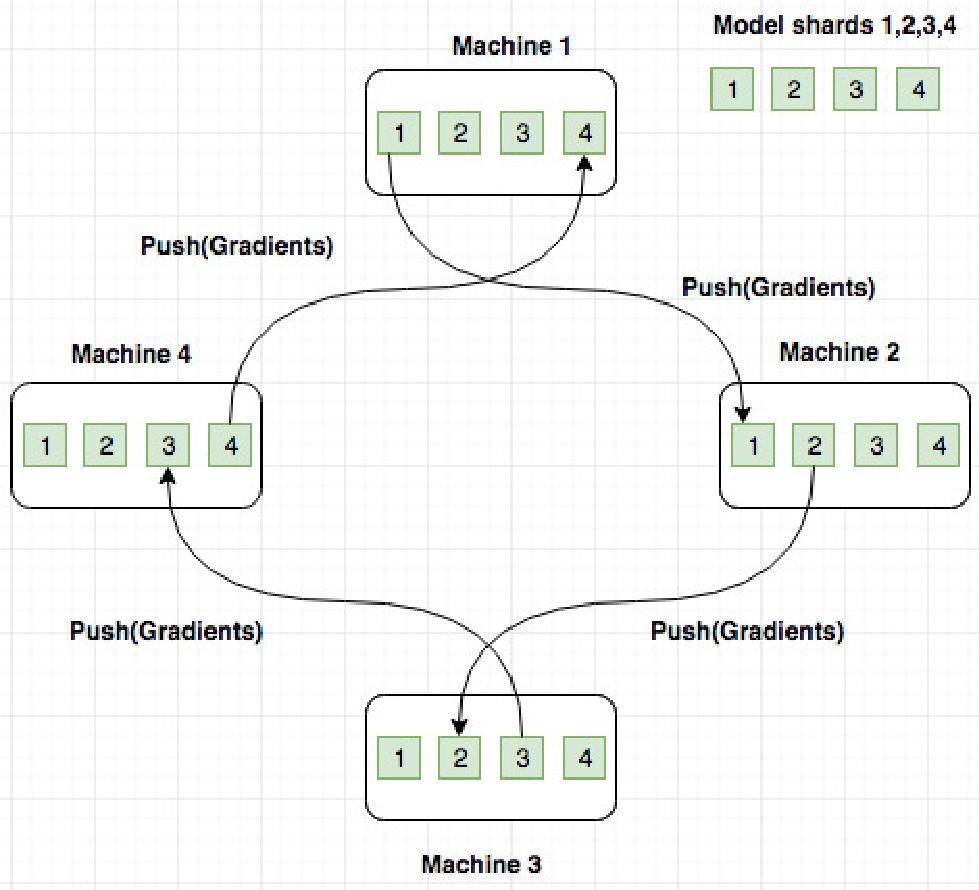}
  \caption{RA Architecture.}
  \label{fig:ring}
\end{figure}

\begin{figure}[htb]
  \includegraphics[width=3.2in]{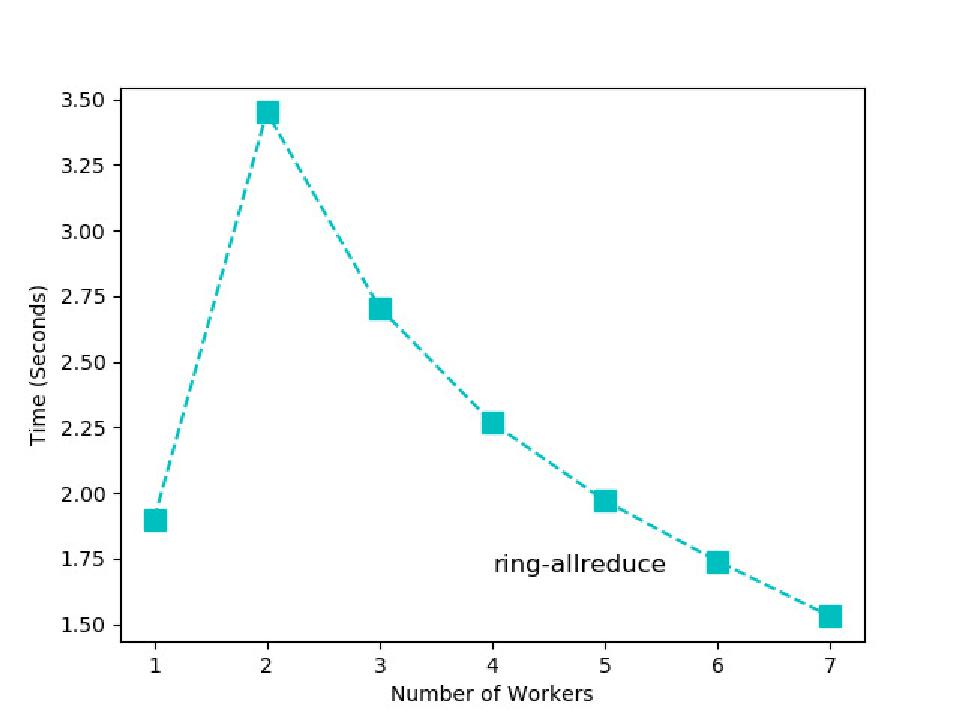}
  \caption{Epoch time of RA.}
  \label{fig:mpilatency}
\end{figure}

\begin{figure}[htb]
  \includegraphics[width=3.2in]{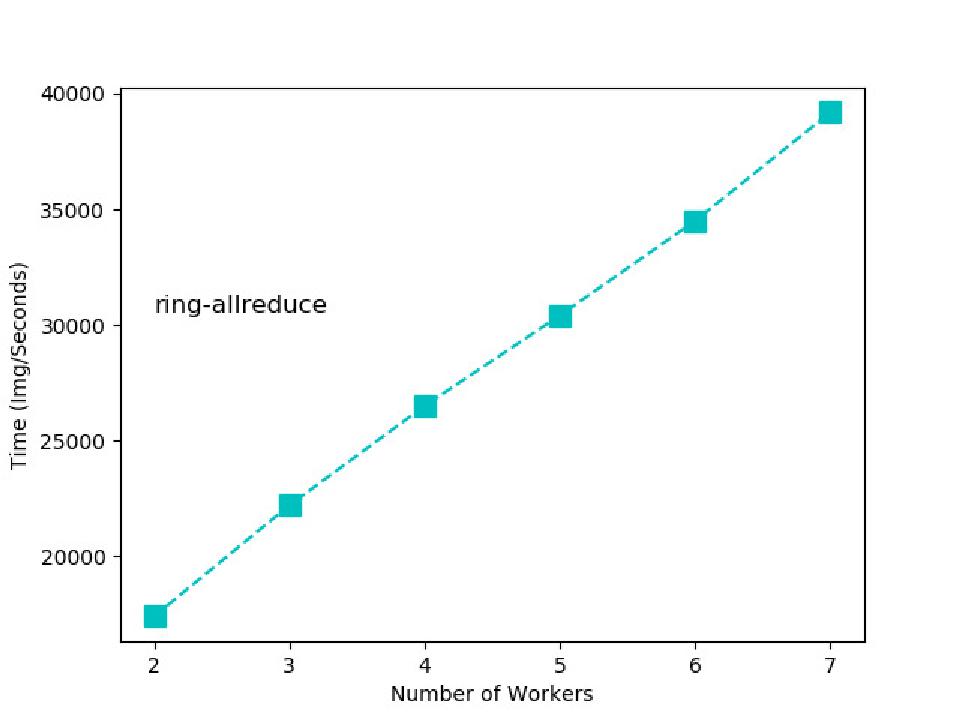}
  \caption{Measured training throughput of RA.}
  \label{fig:mpithroughput}
\end{figure}

In this system architecture, as shown in Figure~\ref{fig:ring}, the first paper was introduced in~\cite{patarasuk2009bandwidth}. Uber Inc adapted the baidu RA algorithm~\cite{baidu-research_2017} and $MPI-Allreduce()$ in its Horovod~\cite{sergeev2018horovod} which is a distributed training framework for TensorFlow. In this model architecture, there is no central server that holds the model and aggregates gradients from workers as PS architecture. Instead, in distributed training, each worker reads its own subset data, calculates its gradients, sends its gradients to its successor node on the ring topology, and receives gradients from its node on the ring topology until all workers have the same values. Based on collected log information, there are many types of communications: negotiate broadcast, broadcast, MPI broadcast, allreduce, MPI allreduce, and negotiate allreduce. Also, MEMCPY IN FUSION BUFFER and MEMCPY OUT FUSION BUFFER are to copy data into and out of the fusion buffer. Each tensor broadcast/reduction in the Horovod involves two major phases. First is the negotiation phase where all workers send a signal to rank 0 that they are ready to broadcast/reduce the given tensor. Then, rank 0 sends a signal to the other workers to start broadcasting/reducing the tensor. Second is the processing phase where the gradients are computed after data loading and preprocessing. Both allreduce and MPI Allreduce are used to average the gradients to single value. The inter-GPU or inter-CPU communication and operation whether on a single network node or across network nodes in implementations of parallel and distributed deep learning training are built on top of MPI and can get benefits from all optimizations that related to MPI such as Open MPI~\cite{Gabriel04openmpi:}. The RA algorithm allows worker nodes to average gradients and send them to all nodes without the need for a PS. The aims of ring reduction operation are to reduce communication overhead that can cause by all-to-one or one-to-all collective communications~\cite{uberjourney}. Horovod also has less amount of code lines for distributed training and increased the scalability comparing to the well-known PS systems. The RA has used distributed data parallelism scheme (see section~\ref{sec:dp}). Every node in the system has a subset of the data $\frac{n}{w}$. The RA~\cite{patarasuk2009bandwidth} is bandwidth-optimal. Technically, every node of $w$ communicates with two of its peers $2*(w-1)$ times. During this communication, a node sends and receives chunks of the data buffer. Every node starts with a local value and ends up with an aggregate global result. In the first $(w-1)$ iterations, each node in a ring topology sends gradients to its successor and receives gradients from its predecessor. Following by a reduction operation that adds up received values to the values in the node's buffer. In the second $(w-1)$ iterations, every node has a sub-final block of data. Finally, all-gather operation transmits a final aggregated block to every other node. The bandwidth is optimal with enough buffer size for storing received messages~\cite{patarasuk2009bandwidth}. RA scales independently of the number of nodes as we find in our experiments~\ref{fig:mpithroughput}. However, RA is limited by the slowest directed communication link between neighbors and available network bandwidth. The latency in the experiment show near steady bandwidth usage that illustrated~\ref{fig:mpilatency}. RA overlaps the computation of gradients at lower layers in a deep neural network with the transmission of gradients at higher layers, which reducing training time.

\begin{equation}
T_{(cpu-ring)} = (epoch* (\frac{\frac{n}{b}}{w}) * (T_{processing}+ update\_time))
\end{equation}

\begin{equation}
T_{(tcp-ring)}  =  \frac{(2*(w-1)*\frac{W}{w})}{Bandwidth}
\end{equation}

\begin{equation}
T_{(total)} = T_{(cpu-ring)} + T_{(tcp-ring)}
\end{equation}

\begin{figure}[htb]
  \includegraphics[width=3.2in]{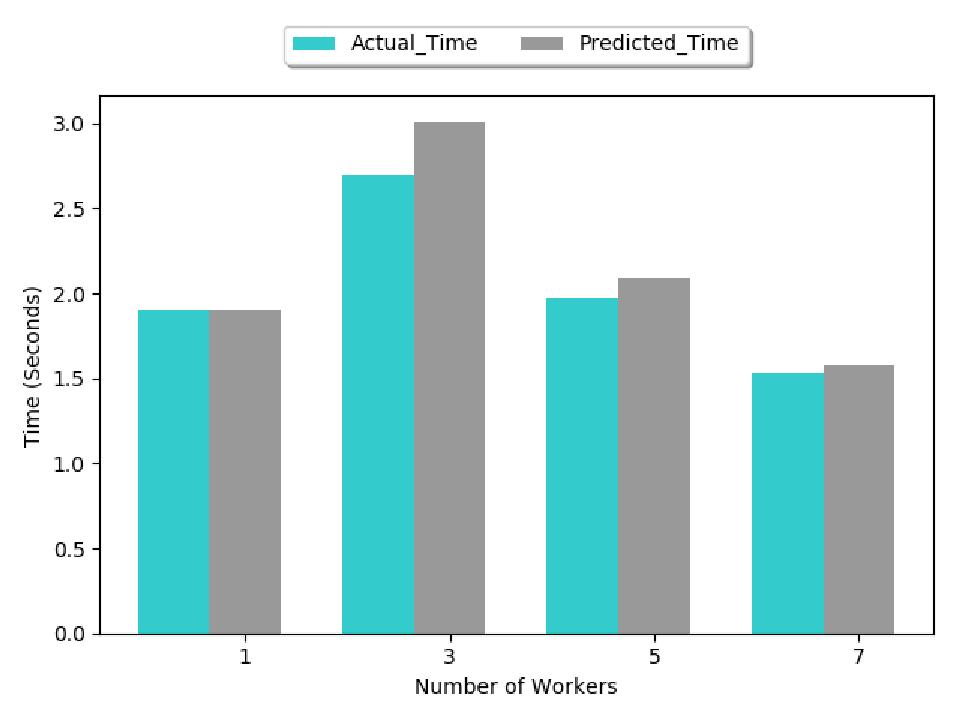}
  \caption{Estimated epoch time for RA.}
  \label{fig:mpimodel}
\end{figure}

\section{Evaluation}
\label{sec:eval}
\subsection{Experimental Environment}

Here, we run a set of experiments in distributed ML system introduced in Section~\ref{sec:back}. In order to provide a quantitative evaluation of 1PS, 2PS, 4PS, RA (Horovod), and P2P Systems, we evaluated the performance of these system architectures with the same basic classification ML tasks. The system performance has two dimensions, latency metric and throughput metric.  All of our experiments were conducted in an Amazon EC2 cloud computing platform using m4.xlarge instances. Each instance contains 4 vCPU powered by Intel Xeon E5-2676 v3 processor and 16GiB RAM. We use the MNIST database of handwritten digits~\cite{MNIST} as our dataset. The MNIST dataset contains 60,000 training samples and 10,000 test samples of handwritten digits (0 to 9). Each digit is normalized and centered in a gray-scale (0 - 255) image with size 28 * 28. Each image consists of 784 pixels that represent the features of the digits. we deployed worker machines from one to seven machines to evaluate and quantify each system throughput and latency. All our ML classification task are written on top of TensorFlow version 1.11.0, an open-source dataflow software library originally release by Google.

\begin{figure}[htb]
  \includegraphics[width=3.2in]{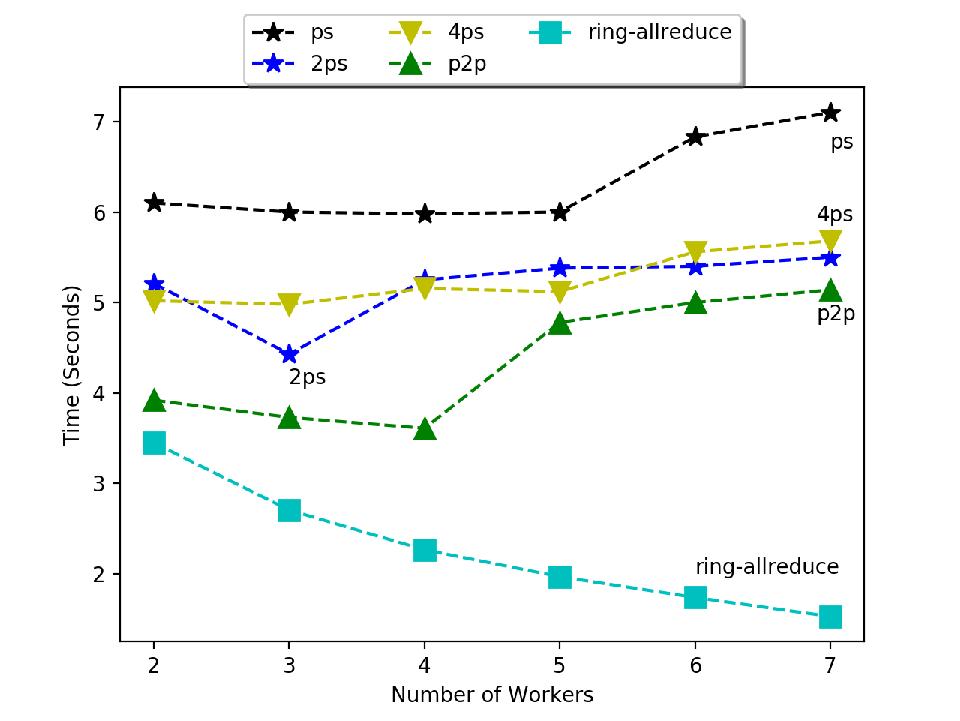}
  \caption{Latency Comparison.}
  \label{fig:clatency}
\end{figure}

\begin{figure}[htb]
  \includegraphics[width=3.2in]{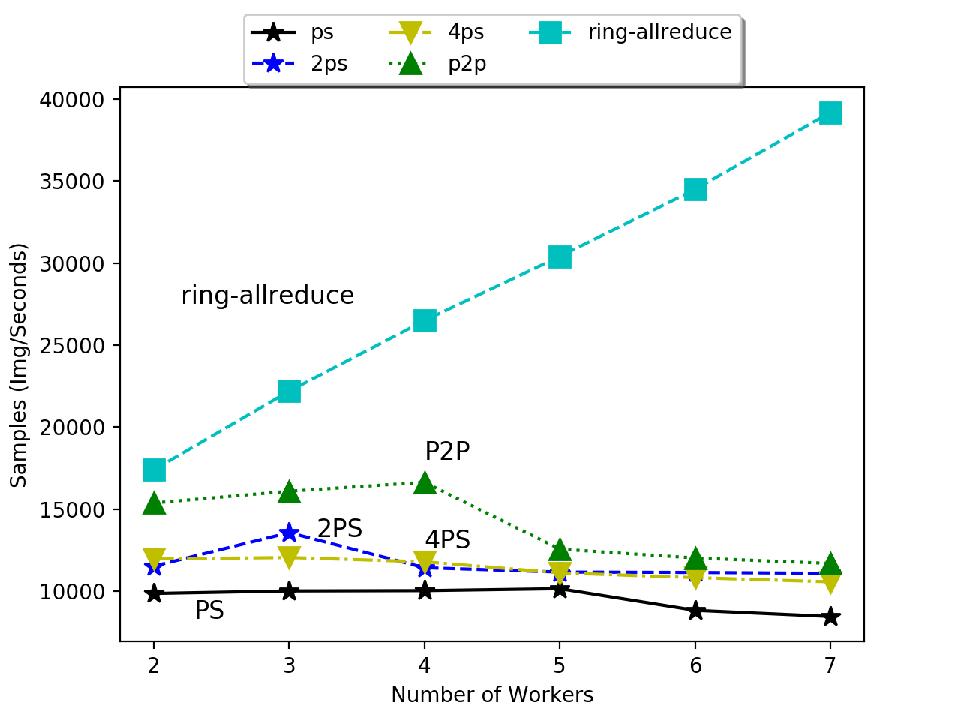}
  \caption{Throughput Comparison.}
  \label{fig:cthroughput}
\end{figure}

\subsection{Experimental Evaluation}

We implemented multilayer neural networks with two hidden layers, and we chose the Softmax activation function as the output layer on three system architectures. We did not include in our study the neural network convergence because we believe that it depends on the neural network architecture and hyper-parameters and it has no dependence of distributed computing framework. For all experiments, we fixed the batch size (Batch size =100). 

\textbf{Parameter Server.} We have three setups of the PS system. The first setup consists of 1PS and workers ranging from one to seven machines located on separated machines. The PS latency at Figure~\ref{fig:pslatency} reduced by adding machines but latency after some point (the fifth machine in our experiment), increased due to all-to-one communication and data overloading which leads to bottleneck on CPU. The nature of a distributed system causes the scalability to degrade after five machines and to increase the epoch time for finishing training cycle. The maximum system throughput as shown in Figure~\ref{fig:psthroughput} was roughly 10150 images per second on five machines. We notice by adding more machine, the throughput is not increasing due to communication bandwidth saturation, and synchronization barriers. The second and third setups consist of two and four PS respectively with workers ranging from one to seven machines located on separated machines. These suffered from network overloaded because more machines communicate with more machines comparing to 1PS. When we optimize the system with 2PS, the maximum throughput was roughly 13570 images per second on three machines as shown in Figure~\ref{fig:2psthroughput}. The latency in Figure~\ref{fig:2pslatency}, on the other hand, increases after three machines. 

\textbf{Peer-to-Peer.} We have the number of servers equal to the number of clients as shown in Figure~\ref{fig:p2p}. In our experiment, we have seven servers and seven clients co-allocated on seven machines. We have noticed some improvement in latency and throughput comparing to 1PS, 2PS, and 4PS as in Figures~\ref{fig:clatency}, and Figure~\ref{fig:cthroughput}. The reason is that the part of model was located on same machine where server located. Also, in this training, we do not need to pull the model from remote machines because it is already updated on the same machine. 

\textbf{Ring Allreduce.} In Figures~\ref{fig:mpilatency}, we noticed that the epoch time reduced sub-linearly with number of machines due to many factors like bandwidth independence from number of nodes, computation and communication overlaps. 

\textbf{Ease of Development} TensorFlow has low-level and high-level API in Python and C++. Back-end TensorFlow was written in C++ while front-end was written in wide language support such as Python and C++. In distributed training, developers have to write and deploy the code on each machine or have to setup the cloud manager. Both ways need expert knowledge to setup and run the network training. In TensorFlow, there are many functions that are available on the framework but because of the updates and frequent new releases with new features deprecation make developers confused. The TensorFlow provides more APIs and primitives than any other ML framework. Debugging is hard but fortunately, we have computational graph visualizations (Tensorboard) that offer the visualization suite for tracking performance and network topology. The TensorFlow has large community support and widely used in many business and labs. Horovod API is different from TensorFlow in many ways such as simplicity of running distributed training. Developers write the code on one machine and that machine will communicate it to every other machine in the system. The Horovod APIs and primitives are not rich comparing to TensorFlow. Horovod has no clear debugging tool and it uses TensorFlow Tensorboard. Horovod has a lack of support and only few business and people are familiar with the Horovod library. In distributed training, I noticed that there is not one right answer for which architecture should be used. Using PS is good when developers have less powerful and not reliable machines such as cluster of CPU. In TensorFlow, PS architecture is well supported and developers will have a large community for help in debugging and suggestions. On the other hand, Horovod is preferable if developers' environments have fast devices such as GPU with strong communication link. 


\section{Related Work}
\label{sec:rw}

Distributed implementation of deep learning algorithms have received much attention in recent years because of its effectiveness in various applications. At present, the usefulness of distributed ML systems such as Tensorflow~\cite{abadi2016tensorflow}, MXNet~\cite{chen2015mxnet}, and Petuum~\cite{xing2015petuum} are recognized in both academia and industry. These open source deep learning frameworks have built on PS architecture. Others systems like~\cite{sergeev2018horovod} has built on RA, or~\cite{li2015malt} has built on a P2P system. Improving performance of these kind of frameworks has a huge impact on computation resources and training time. 

Several works focus on analyzing performance from single system design perspective but for our knowledge, there has not been an in depth comparison study of communication performance with different systems' architectures. Recently, we have published a comparison study of design approaches used in distributed ML platforms such as TensorFlow, MXNet, and Spark~\cite{KUO1}. We focused on system scalability, graph computation speed, fault-tolerance, ease-of-programming, and  resource consumptions to identify the difference between these framework designs and their bottlenecks. 

The performance limits in Apache Spark for distributed ML applications are scalability and compares with high performance computing MPI framework~\cite{Spark}. With some optimization techniques, Spark implementation has performed on learning task better than MPI framework on an equivalent task. These optimization techniques reduced some of training overhead due to language dependency. However, the best performance and alleviation overhead can come from tuning distributed algorithm and distributed system framework properties. A recent work~\cite{hashemi2016performance} focused on analyzing DNNs performance by using CNTK framework. The performance model was to capture the scalability of the system while increasing the computation nodes in small and large clusters. The paper concluded that the CNTK suffers from poor I/O that degraded computation time. Main~\cite{mai2015optimizing} designed called MLNET, a novel communication layer to solve network bottlenecks for distributed ML using tree-based overlays to implement distributed aggregation and multicast and reduce network traffic.


\section{Conclusion}
\label{sec:concl}

In this work, we presented a comparative analysis of communication performance (latency and throughput) for three distributed system architectures.  We found in 1PS, 2PS, 4PS that the throughput fails to increase linearly due to network congestion. We also found that RA achieves better performance due to the efficient use of network bandwidth and overlapping computation and communication. 

We hope our study would help the practitioners for selecting the system architecture and deployment parameters for their training systems. 
Our study can also pave the way for future work to estimate the scalability of distributed DNNs training. A promising direction for future work is to study the trade-offs between network congestion and extra computation. The research question here from the distributed system perspective is to identify which architectures and design elements can facilitate exploring and exploiting these trade-offs.

\section{Acknowledgments}
\label{sec:Acknow}

This project is in part sponsored by the National Science Foundation (NSF) under award number CNS-1527629 and XPS-1533870.

\bibliographystyle{IEEEtran}
\bibliography{ICCCNDatabase}

%
\end{document}